\documentclass{article}
\usepackage{graphicx}
\usepackage{epstopdf}
\usepackage{dialogue}
\usepackage{amsmath}
\usepackage{tcolorbox}
\usepackage{framed}
\usepackage{amsfonts}

\def\noheaderplainsetup{%

    \topmargin=0pt \headheight=0pt \headsep=0pt  
    \oddsidemargin=0pt \evensidemargin=0pt       
    \textheight=8.9truein \textwidth=6.5truein}  

\noheaderplainsetup
\begin{document}

\newcommand{\tasktime}{\textbf{\texttt{TIME}}}
\newcommand{\taskbalance}{\textbf{\texttt{BALANCE}}}
\newcommand{\tasktemp}{\textbf{\texttt{TEMPERATURE}}}
\newcommand{\taskpregnancy}{\textbf{\texttt{PREGNANCY}}}

\newcommand\sbullet[1][.5]{\mathbin{\vcenter{\hbox{\scalebox{#1}{$\bullet$}}}}}

\newcommand{\inp}[1]{{\mathbf Int}^{#1}}
\newcommand{\lit}{\mbox{\bf Literal}}
\newcommand{\por}{\mbox{\bf Port}}
\newcommand{\fn}[1]{\mbox{\bf Tfn}^{#1}}
\newcommand{\str}[1]{\mbox{\bf St}^{#1}}
\newcommand{\rsp}[1]{\mbox{\bf Pl}^{#1}}

\newcommand{\seq}[1]{\langle #1 \rangle}           

\newcommand{\mla}{\mbox{{\Large $\wedge$}}}
\newcommand{\mle}{\mbox{{\Large $\vee$}}}

\newcommand{\pst}{\mbox{\raisebox{-0.01cm}{\scriptsize $\wedge$}\hspace{-4pt}\raisebox{0.16cm}{\tiny $\mid$}\hspace{2pt}}}
\newcommand{\gneg}{\neg}                  
\newcommand{\mli}{\rightarrow}                     
\newcommand{\cla}{\mbox{\large $\forall$}}      
\newcommand{\cle}{\mbox{\large $\exists$}}        
\newcommand{\mld}{\vee}    
\newcommand{\mlc}{\wedge}  
\newcommand{\ade}{\mbox{\Large $\sqcup$}}      
\newcommand{\ada}{\mbox{\Large $\sqcap$}}      
\newcommand{\add}{\sqcup}                      
\newcommand{\adc}{\sqcap}                      

\newcommand{\tlg}{\bot}               
\newcommand{\twg}{\top}               
\newcommand{\st}{\mbox{\raisebox{-0.05cm}{$\circ$}\hspace{-0.13cm}\raisebox{0.16cm}{\tiny $\mid$}\hspace{2pt}}}
\newcommand{\cost}{\mbox{\raisebox{0.12cm}{$\circ$}\hspace{-0.13cm}\raisebox{0.02cm}{\tiny $\mid$}\hspace{2pt}}}
\newcommand{\pcost}{\mbox{\raisebox{0.12cm}{\scriptsize $\vee$}\hspace{-4pt}\raisebox{0.02cm}{\tiny $\mid$}\hspace{2pt}}}

\newtheorem{theoremm}{Theorem}[section]
\newtheorem{thesiss}[theoremm]{Thesis}
\newtheorem{definitionn}[theoremm]{Definition}
\newtheorem{lemmaa}[theoremm]{Lemma}
\newtheorem{propositionn}[theoremm]{Proposition}
\newtheorem{conventionn}[theoremm]{Convention}
\newtheorem{examplee}[theoremm]{Example}
\newtheorem{remarkk}[theoremm]{Remark}
\newtheorem{factt}[theoremm]{Fact}
\newtheorem{exercisee}[theoremm]{Exercise}

\newenvironment{exercise}{\begin{exercisee} \em}{ \end{exercisee}}
\newenvironment{definition}{\begin{definitionn} \em}{ \end{definitionn}}
\newenvironment{theorem}{\begin{theoremm}}{\end{theoremm}}
\newenvironment{lemma}{\begin{lemmaa}}{\end{lemmaa}}
\newenvironment{proposition}{\begin{propositionn} }{\end{propositionn}}
\newenvironment{convention}{\begin{conventionn} \em}{\end{conventionn}}
\newenvironment{remark}{\begin{remarkk} \em}{\end{remarkk}}
\newenvironment{proof}{ {\bf Proof.} }{\  $\Box$ \vspace{.1in} }
\newenvironment{example}{\begin{examplee} \em}{\end{examplee}}
\newenvironment{fact}{\begin{factt}}{\end{factt}}

\newcommand{\chand}{\sqcap} 

\newcommand{\pand}{\wedge} 

\newcommand{\sand}{\hspace{2pt}\mbox{\small \raisebox{0.0cm}{$\bigtriangleup$}}\hspace{2pt}} 

\newcommand{\tand}{\mbox{\hspace{2pt}$\wedge$\hspace{-1.29mm}\raisebox{0.02mm}{\rule{0.13mm}{2mm}}}\hspace{5pt}}    

\newcommand{\chor}{\sqcup} 

\newcommand{\sor}{\hspace{2pt}\mbox{\small \raisebox{0.049cm}{$\bigtriangledown$}}\hspace{2pt}} 

\newcommand{\tor}{\mbox{\hspace{2pt}$\vee$\hspace{-1.29mm}\raisebox{0.1mm}{\rule{0.13mm}{2mm}}\hspace{5pt}}}    

\newcommand{\blall}{\mbox{\large $\forall$}} 

\newcommand{\chall}{\hspace{1pt}\mbox{\Large $\sqcap$}} 

\newcommand{\pall}{\hspace{1pt}\mbox{{\Large $\wedge$}}\hspace{1pt}} 

\newcommand{\sall}{\mbox{\large \raisebox{0.0cm}{$\bigtriangleup$}}} 

\newcommand{\tall}{\mbox{\hspace{1pt}\Large $\wedge$\hspace{-1.84mm}\raisebox{0.02mm}{\rule{0.13mm}{3.0mm}}\hspace{6pt}}}   

\newcommand{\blexists}{\mbox{\large $\exists$}} 

\newcommand{\chexists}{\hspace{1pt}\mbox{\Large $\sqcup$}} 

\newcommand{\pexists}{\hspace{0pt}\mbox{{\Large $\vee$}}\hspace{0pt}} 

\newcommand{\sexists}{\mbox{\large \raisebox{0.07cm}{$\bigtriangledown$}}} 

\newcommand{\texists}{\hspace{1pt}\mbox{\Large $\vee$\hspace{-1.84mm}\raisebox{0.1mm}{\rule{0.13mm}{3.0mm}}\hspace{5pt}}} 

\newcommand{\chimplication}{\sqsupset} 

\newcommand{\pimplication}{\rightarrow} 

\newcommand{\simplication}{\hspace{3pt}\mbox{\Large $\triangleright$}\hspace{3pt}} 

\newcommand{\timplication}{>\hspace{-11pt}-\hspace{2pt}} 

\newcommand{\precurrence}{\hspace{1pt}\mbox{\raisebox{-0.01cm}{\scriptsize $\wedge$}\hspace{-4pt}\raisebox{0.16cm}{\tiny $\mid$}}\hspace{2pt}} 

\newcommand{\srecurrence}{\mbox{\raisebox{-0.07cm}{\scriptsize $-$}\hspace{-5.9pt}\mbox{\raisebox{-0.01cm}{\scriptsize $\wedge$}\hspace{-4pt}\raisebox{0.16cm}{\tiny $\mid$}}}\hspace{2pt}} 

\newcommand{\trecurrence}{\mbox{\raisebox{-0.01cm}{\scriptsize $\wedge$}\hspace{-3.95pt}\raisebox{0.06cm}{\small $\mid$}\hspace{2pt}}}  

\newcommand{\brecurrence}{\mbox{\raisebox{-0.05cm}{$\circ$}\hspace{-0.13cm}\raisebox{3.9pt}{\tiny $\mid$}}\hspace{1.5pt}} 

\newcommand{\coprecurrence}{\hspace{1pt}\mbox{\raisebox{0.12cm}{\scriptsize $\vee$}\hspace{-3.8pt}\raisebox{0.02cm}{\tiny $\mid$}}\hspace{2pt}} 

\newcommand{\cosrecurrence}{\mbox{\raisebox{0.20cm}{\scriptsize $-$}\hspace{-5.9pt}\mbox{\raisebox{0.12cm}{\scriptsize $\vee$}\hspace{-3.8pt}\raisebox{0.02cm}{\tiny $\mid$}}}\hspace{2pt}} 

\newcommand{\cotrecurrence}{\mbox{\raisebox{0.12cm}{\scriptsize $\vee$}\hspace{-3.95pt}\raisebox{0.04cm}{\small $\mid$}\hspace{2pt}}}  

\newcommand{\cobrecurrence}{\mbox{\raisebox{0.12cm}{$\circ$}\hspace{-0.13cm}\raisebox{0.02cm}{\tiny $\mid$}}\hspace{1.5pt}} 

\newcommand{\primplication}{\hspace{2pt}\mbox{\raisebox{0.033cm}{\tiny $>$}\hspace{-0.18cm} \raisebox{-0.043cm}{\large --}}\hspace{2pt}} 

\newcommand{\srimplication}{\hspace{3pt}\mbox{\mbox{\raisebox{0.1pt}{\small $\triangleright$}}\hspace{-4pt}  \raisebox{-0.8pt}{\large --}}\hspace{3pt}} 

\newcommand{\trimplication}{\mbox{\hspace{2pt}\raisebox{0.033cm}{\tiny $>$}\hspace{-0.28cm} \raisebox{-2.4pt}{\LARGE --}\hspace{2pt}}}

\newcommand{\brimplication}{\hspace{3pt}\mbox{$\circ$\hspace{-0.14cm} \raisebox{-0.043cm}{\Large --}}\hspace{3pt}} 

\newcommand{\prepudiation}{\mbox{\raisebox{1.1pt}{\tiny $>$}\hspace{-1.6pt}{\scriptsize $\neg$}}} 

\newcommand{\srepudiation}{\hspace{1pt}\mbox{\raisebox{0.1pt}{\small $\triangleright$}\hspace{-1pt}{\scriptsize $\neg$}}} 

\newcommand{\trepudiation}{\mbox{\raisebox{0.033cm}{\tiny $>$}\hspace{-0.28cm} \raisebox{-0.5pt}{\large -}\hspace{-1pt}{\scriptsize $\neg$}}} 

\newcommand{\brepudiation}{\mbox{$\circ$\hspace{-0.14cm} \hspace{-0.8pt}\raisebox{0.2pt}{\scriptsize $\neg$}}} 

\title{Cirquent Calculus in a Nutshell \footnote{ This paper is based on presentations of cirquent calculus given by the first author at the conferences {\em Smirnov Readings} (Moscow, June 21, 2019) and {\em Logic, Quantum Computing, and Artificial Intelligence} (Online, July 3, 2021). }} 

\author{Giorgi Japaridze$^{+}$,  Bikal Lamichhane$^{+}$ 
\\  \\ $^{+}$Department of Computing Sciences, Villanova University, 
\\ 800 Lancaster Avenue, Villanova, PA 19085, USA\\
}
\date{}
\maketitle

\begin{abstract} This paper is a brief and informal presentation of  \emph{cirquent calculus}, a novel proof system for resource-conscious logics. As such, it is a refinement of sequent calculus with mechanisms that allow to explicitly account for the possibility of sharing of subexpressions/subresources between different expressions/resources. This is achieved by dealing with circuit-style constructs, termed \emph{cirquents},  instead of formulas, sequents or other tree-like structures. The approach  exhibits greater expressiveness, flexibility and efficiency compared to the more traditional  proof-theoretic approaches. The need for substantially new deductive tools that could overcome limitations of sequent calculus while dealing with resource logics surfaced with the birth of \emph{computability logic}, a game-semantically conceived logic of computational resources and tasks, acting as a formal theory of computability in the same sense as classical logic is a formal theory of truth.Cirquent calculus offers elegant axiomatizations for certain basic fragments of computability logic that have been shown to be inherently unaxiomatizable in sequent calculus or other traditional systems.  
\end{abstract}

\section{Introduction}
\textbf{Cirquent calculus}, introduced in \cite{Cirq} and further developed in  \cite{bauerLMCS}, \cite{Japdeep}-\cite{cl16}, \cite{XuIGPL}-\cite{XuLast}, is a novel proof-theoretic approach, more general than traditional approaches such as sequent calculus, Hilbert-style systems, natural deduction, etc. --- general in the sense that such systems can always be translated into cirquent calculus but not vice versa.  The main distinguishing feature of cirquent calculus  is that it allows sharing of subcomponents between different components of its expressions. Instead of formulas, it operates with circuit-style constructs termed {\em cirquents}.  The word `cirquent' is a portmanteau of `circuit' and  `sequent': as we're going to see, cirquents are in fact generalized sequents.

To see why circuits are chosen over formulas, one can ask the following rhetorical question:  what is a more natural representation of Boolean functions, formulas or circuits? 

\begin{center} 

\begin{picture}(300,90)
\put(20,50){$(A\mlc B)\mld(A \mlc C) \mld (B \mlc C)$}
\put(180,75){$A$}
\put(208,75){$B$}
\put(238,75){$C$}
\put(184,78){\circle{15}}
\put(213,78){\circle{15}}
\put(242,78){\circle{15}}

\put(184,59){\line(5,2){28}}
\put(184,59){\line(0,1){11}}
\put(213,59){\line(-5,2){28}}
\put(213,59){\line(5,2){29}}
\put(242,59){\line(-5,2){30}}
\put(242,59){\line(0,1){12}}
\put(184,51){\circle{15}}
\put(213,51){\circle{15}}
\put(242,51){\circle{15}}
\put(213,25){\circle{15}}
\put(181,48){$\mlc$}
\put(210,48){$\mlc$}
\put(239,48){$\mlc$}
\put(210,22){$\mld$}
\put(213,33){\line(0,1){10}}
\put(213,33){\line(5,2){27}}
\put(213,33){\line(-5,2){27}}
\put(120,0){\bf Figure 1}
\end{picture}
\end{center}
 \vspace{5pt}
\noindent In Figure 1  we see  a formula and a circuit for the same Boolean function. Notice the sharing that is taking place in the circuit. For instance,  input  $B$ is shared between the first and the third conjunctive gates. Formulas, on the other hand, have no sharing mechanisms, so  $B$   had to be written twice. Due to  the sharing of subcomponents, circuits are generally exponentially more compact than formulas. Even though this fact might not be appreciated by just looking at  Figure 1, with increase in the complexity of Boolean functions it starts becoming more evident. The compactness of circuits is the reason why computer hardware is based on circuits rather than formulas or other structures that take no advantage of sharing.

Cirquents are more general than Boolean circuits: every Boolean circuit is a cirquent but not vice versa. Out of the two cirquents shown  below, 

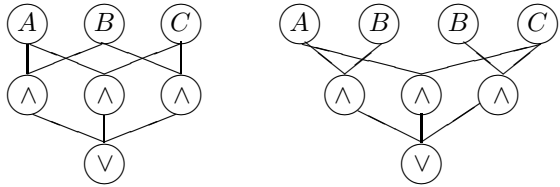
\begin{center}
\begin{picture}(180,65)

\put(0,55){$A$}
\put(29,55){$B$}
\put(58,55){$C$}

\put(4,58){\circle{15}}
\put(33,58){\circle{15}}
\put(62,58){\circle{15}}

\put(4,39){\line(5,2){28}}
\put(4,39){\line(0,1){11}}
\put(33,39){\line(-5,2){28}}
\put(33,39){\line(5,2){29}}
\put(62,39){\line(-5,2){30}}
\put(62,39){\line(0,1){12}}

\put(4,31){\circle{15}}
\put(33,31){\circle{15}}
\put(62,31){\circle{15}}
\put(33,5){\circle{15}}
\put(1,28){$\mlc$}
\put(30,28){$\mlc$}
\put(59,28){$\mlc$}
\put(30,2){$\mld$}
\put(33,13){\line(0,1){10}}
\put(33,13){\line(5,2){27}}
\put(33,13){\line(-5,2){27}}

\put(103,55){$A$}
\put(133,55){$B$}
\put(163,55){$B$}
\put(193,55){$C$}

\put(107,58){\circle{15}}
\put(137,58){\circle{15}}
\put(167,58){\circle{15}}
\put(197,58){\circle{15}}

\put(124,39){\line(-5,4){15}}
\put(124,39){\line(5,4){14}}
\put(153,39){\line(-4,1){45}}
\put(153,39){\line(4,1){45}}
\put(184,39){\line(-5,4){15}}
\put(184,39){\line(5,4){14}}

\put(124,31){\circle{15}}
\put(153,31){\circle{15}}
\put(182,31){\circle{15}}
\put(153,5){\circle{15}}
\put(120,28){$\mlc$}
\put(150,28){$\mlc$}
\put(180,28){$\mlc$}
\put(150,2){$\mld$}

\put(153,13){\line(-2,1){24}}
\put(153,13){\line(0,1){11}}
\put(153,13){\line(3,2){22}}
\end{picture}
\end{center}

\noindent only the one on the left is a Boolean circuit in the proper sense. The one on the right is not how Boolean circuits are normally drawn as it has two copies of $B$.  From the classical point of view,  writing (the ``same'' input) $B$ twice makes makes no sense. On the other hand, from the point of view of cirquent calculus, the above two are simply diffferent cirquents, not only syntactically but also semantically. 

Additionally, cirquents are more general than Boolean circuits in that, when dealing with certain non-classical logics, the former may have non-Boolean gates just as well.  Such cirquents can be found in \cite{fromto,taming1,taming2,cl16,cl17,XuIf,XuLast}.

The main motivations  for developing cirquent calculus have been the following:
\begin{itemize}
    \item \textbf{To axiomatize computability logic}: Computability logic, which we shall henceforth abbreviate as {\bf CoL}, is a semantically conceived logic of computational tasks and resources,  originally introduced in \cite{Jap03} and subsequently developed in a long series of publications (see \cite{fundamentals} for a survey). For quite some time, CoL had stubbornly resisted all axiomatization attempts within the framework of traditional proof systems. Eventually it was simply proven in  \cite{anupam} that axiomatizing even the most basic fragment of CoL in traditional systems was impossible in principle. As an nontraditional approach, cirquent calculus comes in and breaks the ice, offering an adequate proof theory for the otherwise untameable CoL.
    
    \item \textbf{To have a more adequate logic of resources than linear logic claims to be}: Here “resources” are meant in a more general/abstract sense than just computational resources studied in CoL. Resource sharing is a ubiquitous phenomenon, yet linear logic, unlike cirquent calculus, lacks mechanisms to account for it. Shame!
    
    \item \textbf{To have more efficient proof systems for classical logic (and beyond) than the traditional ones}: As illustrated in \cite{Japdeep},  cirquent  calculus achieves an exponential speedup of analytic proofs over cut-free sequent calculus or other analytic proof systems. Among the tautology classes enjoying this sort of a proof speedup is the well-studied Pigeonhole Principle.

\end{itemize}
    
\section{Infocomputational tasks and operations on them}

Computability logic has a whole zoo of logical opeartors, listed  below. There are several sorts of disjunctions, conjunctions, quantifiers, and more. Among them, in this article we will only be dealing with $\neg$, $\mlc$ and $\mld$.

\begin{itemize}
  \item Negation:  $\neg$.\vspace{-5pt}
  \item Conjunctions:  $\pand$ (parallel);    $\chand$ (choice);   $\sand$ (sequential);  $\tand$ (toggling).\vspace{-5pt}
  \item Disjunctions:       $\mld$ (parallel);     $\chor$ (choice);   $\sor$ (sequential);    $\tor$ (toggling).\vspace{-5pt}
  \item Implications:    $\pimplication$ (parallel);     $\chimplication$ (choice);    $\simplication$ (sequential);    $\timplication$ (toggling).\vspace{-5pt}
  \item Universal quantifiers:   $\blall$ (blind);  $\pall$ (parallel);     $\chall$ (choice);   $\sall$ (sequential);    $\tall$ (toggling).  \vspace{-5pt}    
  \item Existential quantifiers:  $\blexists$ (blind);    $\pexists$  (parallel);     $\chexists $ (choice);   $ \sexists$ (sequential);   $\texists$ (toggling).\vspace{-5pt}      
  \item Recurrences:  $\brecurrence$ (branching);    $\precurrence$  (parallel);      $\srecurrence$ (sequential);   $\trecurrence$ (toggling).\vspace{-5pt}    
  \item Corecurrences:    $\cobrecurrence$ (branching);    $\coprecurrence$  (parallel);      $\cosrecurrence$ (sequential);   $\cotrecurrence$ (toggling).\vspace{-5pt}
  \item Rimplications:       $\brimplication$ (branching);    $\primplication$ (parallel);      $\srimplication$ (sequential);   
 $\trimplication$ (toggling).\vspace{-5pt}   
  \item Repudiations:         $\brepudiation$ (branching);    $\prepudiation$ (parallel);      $\srepudiation$ (sequential);    $\trepudiation$ (toggling). 
\end{itemize}

In CoL, formulas stand for {\bf infocomputational} (informational/computational) tasks, and logical operators for operations on such entities. We do not differentiate betweem ``informational'' and ``computational'' here because, after all, the purpose of every computation is to provide information, whether the latter is immediately fetched from some database or is obtained as a result of running some sophisticated algorithms. Mathematically, infocomputational tasks  are defined as games between two agents, called  {\bf Provider} and {\bf Consumer}  (more often referred to as the machine and the environment, respectively). 

Intuitively, such games can be seen as dialogues consisting of questions and answers. 
For instance, the  task performed by the watch (Provider) that you (Consumer) are wearing---call that task $\tasktime$---is to tell you the current time  whenever you ``ask'' (well, look at) it. A ``dialogue'' between you and your watch takes the following form:
\begin{dialogue}
\speak{Consumer} What time is it now?
\speak{Provider} It is 1:30 PM.
\end{dialogue}

\noindent Seeing the above as a game, it is a two-move game where the first move,``What time is it now?'',  is by Consumer and the second move,  ``It is 1:30 PM'', by Provider.  Provider is considered to have accomplished the task, i.e., won the current run of the game, if it correctly answered the question, i.e., if it is indeed 1:30 PM. Otherwise it is Consumer who wins. As an aside, this is one case where you do not want to be the winner. If your watch lies to you, then you may lose a
 more important ``game'', such as catching the plane. Generally, the same is the case whenever you ``play'' agains an agent meant to be a tool for you to use and rely on.

Most tasks, however, will be more interactive than what we just saw, with multiple questions and answers by both parties.  Consider the dialogue below.

\begin{dialogue}
\speak{Consumer} What is my balance?
\speak{Provider}  What is your account number?
\speak{Consumer} My account number is 333444555.
\speak{Provider} What is your password?
\speak{Consumer} My password is ``password123".
\speak{Provider} Your balance is \$12000.
\end{dialogue}
\noindent It is for the task $\taskbalance$ performed by an automated teller machine (ATM) of a bank. Consumer inquires about his/her balance. But the ATM (=Provider) releases the balance details only after asking (counter)questions and receiving answers regarding the account number and pasword.  Here, Provider has accomplished the task if Consumer's balance is indeed \$12000, or if Consumer has failed to provide the right account number and password.

Below is an informal explanation of the meanings of the three game operations on which the present article is focused. Formal definitions of all operations can be looked up in \cite{fundamentals}. 

\textbf{Negation ($\neg$)} interchanges the roles of the two agents: Provider’s questions or answers become those of Consumer, and vice versa. For example, as in $\tasktime$ Consumer asks the time and Provider answers, in $\neg\tasktime$ it is Provider who asks such a question, and it is Consumer’s obligation to answer. A naive reader may ask here: ``But wait a second, why do we continue calling that agent ``Provider'' if it no longer provides any information but just the opposite, it consumes information?'' Well,  ``Provider'' and “Consumer” are  just names that can be assigned to two agents as one pleases. After all, for either agent, most taks would naturally involve both providing and consuming information, as was the case with the above task   $\taskbalance$. Speaking of tasks, they are symmetric to resources: what is a task for Provider is a resource for Consumer, and vice versa. So, CoL is a logic of infocomputational resources just as much as it is a logic of infocomputational tasks. 

The \textbf{parallel conjunction $A \mlc B$} of tasks $A$ and $B$ is the task where, in order to accomplish it, Provider needs to accomplish both $A$ and $B$. In other words, this is a game on two “boards”, where Provider wins 
if it wins on both boards. From Consumer’s perspective, having the resource $A \mlc B$ means having both $A$ and $B$.  For example, if you have a  functioning watch and a functioning thermometer, the resource you possess is $\tasktime\mlc\tasktemp$. It allows you to simultaneously know both the current time and the current temperature. 

\textbf{Parallel disjunction $A \mld B$} can be formally defined as $\neg(\neg A \mld \neg B)$. This is a two-board game just like $A \mlc B$, with the difference that here provider is only obligated to win in at least one of the two components rather than both. Intuitively, possessing the resource $\tasktime  \mld \tasktemp$ corresponds to the situation in which you have a watch and a thermometer,  where one of them might be malfunctioning but you don’t know which one. So, if they tell you  “8 AM” and “15\textdegree C’’, you will only come to know that either the time is 8 AM  or the temperature is 15\textdegree C, without otherwise being sure which one is true. 

\section{Semantics}
 CoL and the corresponding cirquent calculus have have their own set of differences with classical and linear logic. For instance, in classical logic we have $A = A \mld A$. However, in CoL, $A \neq A \mld A!$. To get a feel for this, let us compare the resources $\tasktime$ and $\tasktime\mld\tasktime$.  Intuitively, $\tasktime$ corresponds to a situation where you have a single watch, and a reliable one. $\tasktime \mld \tasktime$, on the other hand, corresponds to a situation where you have two watches, and you only know that at least one of them is reliable, even though you cannot tell which one. Obviously $\tasktime$ is more helpful than $\tasktime\mld\tasktime$. That is, as a resource, the former is stronger than the latter.

A task is defined to be {\bf valid} if (a smart) Provider has a strategy to always accomplish it without any physical or informational resources. That is, a valid task is one that can be accomplished by a Provider who knows nothing   else but logic. Of course, here the word ``defined'' should be taken in quotation marks: this presentation deliberately avoids any strict definitions, relying on informal explanations instead. Formal definitions, of course, do exists for all concepts introduced here. Again, they, for instance, can be found in \cite{fundamentals}. 

Any task of the form  $\neg A \mld A$ is valid. E.g., Provider's winning strategy for  $\neg\tasktime\mld\tasktime$ is to wait for Consumer to ask ``what time is it?" in $\tasktime$, and ask the same (counter)question in $\neg \tasktime$; then, whatever Consumer answers, repeat the same answer in $\tasktime$. A few other examples of valid principles are:
\begin{itemize}
  \item $A \mld B \mld \neg A$
  \item $ ((\neg A \mld \neg B) \mlc (\neg C \mld \neg D)) \mld ((A \mld C) \mlc (B \mld D)) $
  \item $ ((\neg A \mld \neg A) \mlc (\neg A \mld \neg A)) \mld ((A \mld A) \mlc (A \mld A)) $
\end{itemize}

Not all classical tautologies are valid though. For instance, $\neg A \mld (A \mlc A)$ is not validated by the semantics of CoL. Let $\taskpregnancy$ be the task of telling any physically present woman's pregnancy status. This is the resource provided by a disposable pregnancy kit. Is there a strategy for accomplishing the task given below?

$$
\underset{\#1}{\neg \taskpregnancy}
\mld (\underset{\#2}{ \taskpregnancy} \mlc \underset{\#3}{ \taskpregnancy})
$$

As a matter of fact, there isn't. Let's try it out.
\begin{dialogue}
\speak{Consumer in \#2:} Is Martha pregnant?
\speak{Provider in \#1:} Is Martha pregnant?
\speak{Consumer in \#1:} Yes, Martha is pregnant.
\speak{Provider in \#2:} Yes, Martha is pregnant.
\speak{Consumer in \#3:} Is Dorothy also pregnant?
\speak{Provider:} ... stuck!
\end{dialogue}

\noindent For the first question posed by Consumer, Provider could utilize \#1 to obtain the correct answer. However, when Consumer poses the question again for Dorothy, Provider gets stuck as it cannot  reutilize \#1 to  answer the question.

Cirquents can be seen as operations on tasks/resources (applied to their atoms/inputs). As such, they are not always expressible in terms of $\neg , \mlc ,\mld$ or other primitives. For instance, the cirquent of Figure 1 is a 3-ary operation, accomplished by Provider iff at least two out of the three components, $A,B,C$ are accomplished. There is no way to write it as a formula. A naive attempt  could try to express it as $(A \mlc B) \mld (A \mlc C) \mld (B \mlc C)$, but it is far from adequate, as it suggests the presence of two copies of the task/resource  $A$, two copies of $B$ and two copies of $C$. In game-semantical terms, the cirquent of Figure 1 is a game on three boards. However, $(A \mlc B) \mld (A \mlc C) \mld (B \mlc C)$ suggests that this is a game on six boards, even if the {\em types} (but not necessarily runs) of the games played on boards $\#1$ and $\#3$ coincide, and so do the types of the games played on boards $\#2$ and $\#5$, or boards $\#4$ and $\#6$.  

The concept of validity now naturally extends from formulas to cirquents: a cirquent is valid iff (the task expressed by) it can always be accomplished by a Provider without any physical or extra-logical informational resources.

\section{Classical and linear logics from the perspective of cirquent calculus}
To understand how ciruent calculus relates to classical logic and linear logic, let us consider the formula \[(A \mlc B) \mld (B \mlc C)\] and ask which of the following two cirquents is a (more) adequate or reasonable representation of its   meaning: 

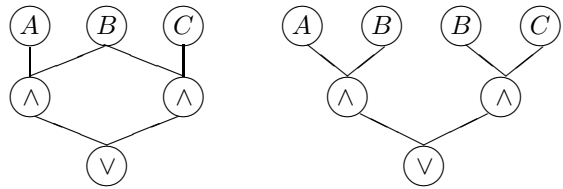
\begin{center}
\begin{picture}(269,76)

\put(36,60){$A$} \put(65,60){$B$} \put(94,60){$C$} \put(40,63){\circle{15}} \put(69,63){\circle{15}} \put(98,63){\circle{15}} \put(40,44){\line(5,2){28}} \put(40,44){\line(0,1){11}} \put(98,44){\line(-5,2){30}} \put(98,44){\line(0,1){12}} \put(40,36){\circle{15}} \put(98,36){\circle{15}} \put(69,10){\circle{15}} \put(37,33){$\mlc$} 
\put(95,33){$\mlc$} \put(66,7){$\mld$} \put(69,18){\line(5,2){27}} \put(69,18){\line(-5,2){27}}

\put(139,60){$A$} \put(169,60){$B$} \put(199,60){$B$} \put(229,60){$C$} \put(143,63){\circle{15}} \put(173,63){\circle{15}} \put(203,63){\circle{15}} \put(233,63){\circle{15}} \put(160,44){\line(-5,4){15}} \put(160,44){\line(5,4){14}} \put(220,44){\line(-5,4){15}} \put(220,44){\line(5,4){14}} \put(160,36){\circle{15}} \put(218,36){\circle{15}} \put(189,10){\circle{15}} \put(156,33){$\mlc$} \put(216,33){$\mlc$} \put(186,7){$\mld$} \put(189,18){\line(-2,1){24}} \put(189,18){\line(2,1){24}}

\end{picture}
\end{center}
The answer depends on who you ask. From the point of view of classical logic, it is the cirquent  on the left: even though in the formula we see $B$ twice, those two occurrences are semantically the same in every relevant sense, so why not combine them together. On the other hand, a linear logician would say that the meaning of the formula is captured by the cirquent on the right: the formula has two copies of $B$, and each copy stands for a separate resource, even if both resources have the same type $B$.  

From the point of view of cirquent calculus, classical logic and linear logic are two imperfect extremes, whose resource-semantical perspectives on logical expressions can be schematically characterized as follows:
\begin{description} 
\item[Classical logic:] {\em Everything is shared} that can be shared. Namely, each occurrence of the same subcomponent stands for the same ``copy'' of the same resource.
\item[Linear logic:] {\em Nothing is shared} at all. Namely, each occurrence of the same subcomponent stands for a different ``copy'' of the same resource.
\end{description}
But how about mixed cases, where some same-type resources may be shared between different subresources and some not, as in the cirquent below where the two conjunctions share both $A$ and $C$ but each one has its own $B$? Only cirquent calculus makes it possible to account for such cases. 
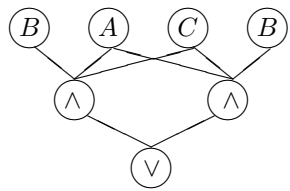
\begin{center}
\begin{picture}(233,76)

\put(73,60){$B$}
\put(103,60){$A$}
\put(133,60){$C$}
\put(163,60){$B$}

\put(77,63){\circle{15}}
\put(107,63){\circle{15}}
\put(137,63){\circle{15}}
\put(167,63){\circle{15}}

\put(94,44){\line(-5,4){15}}
\put(94,44){\line(5,4){14}}
\put(94,44){\line(4,1){45}}

\put(154,44){\line(-4,1){45}}
\put(154,44){\line(-5,4){15}}
\put(154,44){\line(5,4){14}}

\put(94,36){\circle{15}}
\put(152,36){\circle{15}}
\put(123,10){\circle{15}}

\put(90,33){$\mlc$}
\put(150,33){$\mlc$}
\put(120,7){$\mld$}

\put(123,18){\line(-2,1){24}}
\put(123,18){\line(2,1){24}}

\end{picture}
\end{center}

\section{Cirquent calculus vs. sequent calculus}

A number of cirquent calculus systems,  of various degrees of expressiveness, have been elaborated in \cite{Cirq}-\cite{cl17}, \cite{XuIGPL}-\cite{XuLast}.   The historically earliest,  simplest and most basic among them is the system \textbf{CL5} introduced in \cite{Cirq}, and this is the only cirquent-calculus system that we are going to look at in this article. 

\textbf{CL5} operates with special---{\em shallow}---sorts of cirquents. Every such cirquent is of height 2, having a conjunctive root with disjunctive children and ($\neg,\mlc,\mld$)-formulas as grandchildren.  Quite simply, a shallow cirquent is a conjunction of disjunctions of formulas. With $F,G,H$ standing for some arbitrary (not necessarily atomic) formulas, below is an example of a shallow cirquent: 
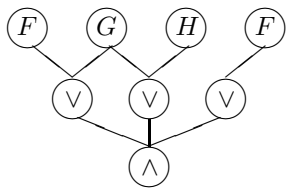
\begin{center} \begin{picture}(150,78)

\put(19,60){$F$}
\put(49,60){$G$}
\put(79,60){$H$}
\put(109,60){$F$}
\put(23,63){\circle{15}}
\put(53,63){\circle{15}}
\put(83,63){\circle{15}}
\put(113,63){\circle{15}}

\put(40,44){\line(-5,4){15}}
\put(69,44){\line(-5,4){15}}
\put(69,44){\line(5,4){15}}
\put(40,44){\line(5,4){15}}
\put(98,44){\line(5,4){15}}
\put(40,36){\circle{15}}
\put(69,36){\circle{15}}
\put(98,36){\circle{15}}
\put(69,10){\circle{15}}
\put(37,33){$\mld$}
\put(66,33){$\mld$}
\put(95,33){$\mld$}
\put(66,7){$\mlc$}
\put(69,18){\line(0,1){10}}
\put(69,18){\line(5,2){27}}
\put(69,18){\line(-5,2){27}}

\end{picture}
\end{center} 

For brevity, in \textbf{CL5} we write such a cirquent without showing (the always-conjunctive) root, and with bullets ($\bullet$)---called {\bf groups}---instead of disjunctive gates, as follows:
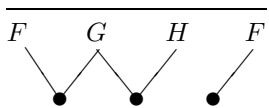
\begin{center} \begin{picture}(118,50)

\put(0,42){\line(1,0){98}}
\put(0,30){$F$}
\put(30,30){$G$}
\put(60,30){$H$}
\put(90,30){$F$}

\put(20,8){\line(-2,3){13}}
\put(49,8){\line(-4,5){15}}
\put(49,8){\line(4,5){15}}
\put(20,8){\line(4,5){15}}
\put(78,8){\line(4,5){15}}
\put(20,8){\circle*{5}}
\put(49,8){\circle*{5}}
\put(78,8){\circle*{5}}
\end{picture}
\end{center} 

\noindent A group is said to {\bf contain} the formulas to which it is connected with an arc; the horizontal line on the top is to show that this is one cirquent rather than two as it may be otherwise perceived. A formula $F$ in isolation is understood as the cirquent with a single group that contains just $F$:
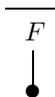
\begin{center} \begin{picture}(30,47)

\put(0,39){\line(1,0){20}}
\put(7,27){$F$}
\put(10,8){\line(0,1){15}}
\put(10,8){\circle*{5}}
\end{picture}
\end{center}

Proof-theoretically, a cirquent can be thought of as a collection of leaf sequents from some (one-sided) sequent-calculus proof tree, where some sequents may share some formulas.
In the following illustration, on the right we see a (fragment of a) sequent-calculus proof tree where the formula $G$ is {\em implicitly} assumed to be shared between the two leaf sequents $\langle F,G\rangle$ and $\langle G,H\rangle$;  on the other hand, the formula $F$, while also occuring in two leaf sequents, is not assumed to be shared between them. The corresponding cirquent on the left, whose three gropus represent the three leaves of the proof tree, makes these implicit assumptions on the presence and absence of sharing explicit.

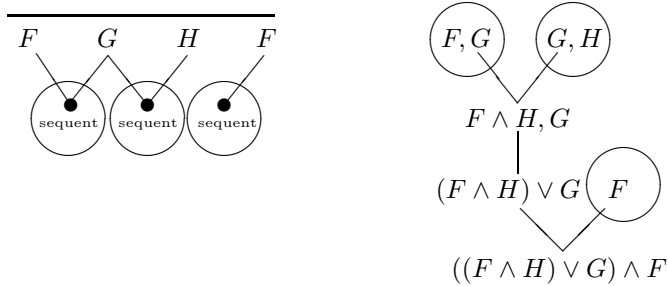
\begin{center} \begin{picture}(236,112)
\put(-4,104){\line(1,0){101}}
\put(0,92){$F$}
\put(30,92){$G$}
\put(60,92){$H$}
\put(90,92){$F$}

\put(20,70){\line(-2,3){13}}
\put(19,70){\line(4,5){15}}

\put(49,70){\line(-4,5){15}}
\put(49,70){\line(4,5){15}}

\put(78,70){\line(4,5){15}}

\put(20,70){\circle*{5}}
\put(49,70){\circle*{5}}
\put(78,70){\circle*{5}}

\put(8,62){\tiny sequent}
\put(38,62){\tiny sequent}
\put(68,62){\tiny sequent}

\put(19,65){\circle{26}}
\put(49,65){\circle{26}}
\put(78,65){\circle{26}}

\put(170,95){\circle{26}}
\put(210,95){\circle{26}}
\put(229,40){\circle{26}}

\put(160,92){$F, G$}
\put(200,92){$G, H$}

\put(189,71){\line(-4,5){15}}
\put(189,71){\line(4,5){15}}

\put(169,62){$F \mlc H, G$}
\put(189,60){\line(0,-1){16}}

\put(158,35){$(F \mlc H) \mld G$}
\put(223,35){$F$}

\put(206,15){\line(-1,1){16}}
\put(206,15){\line(1,1){16}}

\put(164, 5){$((F \mlc H) \mld G) \mlc F$}

\end{picture}
\end{center}

The inference rules of {\bf CL5} will only be explained in the following section. Nevertheless, for preliminary intuitive insights into the superiority of cirquent calculus over sequent calcus, let us still try to compare a purported cirquent-calculus proof (whatever that precisely means)---or rather a fragment of such a proof--- of the formula $((F \mlc H) \mld G) \mlc F$ with a corresponding fragment of a sequent-calculus proof of the same formula, shown in Figure 2. 

\begin{center} \begin{picture}(236,200)
\put(0,180){\line(1,0){98}}

\put(0,168){$F$}
\put(30,168){$G$}
\put(60,168){$H$}
\put(90,168){$F$}

\put(20,146){\line(-2,3){13}}
\put(20,146){\line(4,5){15}}
\put(49,146){\line(-4,5){15}}
\put(49,146){\line(4,5){15}}
\put(78,146){\line(4,5){15}}
\put(20,146){\circle*{5}}
\put(49,146){\circle*{5}}
\put(78,146){\circle*{5}}

\put(0,140){\line(1,0){98}}

\put(0,128){$F \mlc H $}
\put(45,128){$G$}
\put(90,128){$F$}

\put(35,106){\line(-4,5){15}}
\put(35,106){\line(4,5){15}}
\put(78,106){\line(4,5){15}}
\put(35,106){\circle*{5}}
\put(78,106){\circle*{5}}

\put(0,100){\line(1,0){98}}

\put(35,66){\line(0,1){18}}
\put(78,66){\line(4,5){15}}
\put(35,66){\circle*{5}}
\put(78,66){\circle*{5}}

\put(0,88){$(F \mlc H) \mld G $}
\put(90,88){$F$}

\put(0,60){\line(1,0){98}}

\put(5,48){$((F \mlc H) \mld G) \mlc F $}
\put(49,23){\line(0,1){20}}
\put(49,23){\circle*{5}}

\put(0,188){cirquent-calculus proof}

\put(160,168){$F, G$}
\put(200,168){$G, H$}

\put(189,138){\line(-4,5){20}}
\put(189,138){\line(4,5){20}}

\put(166,128){$F \mlc H, G$}

\put(189,122){\line(0,-1){23}}

\put(157,88){$(F \mlc H) \mld G$}
\put(238,88){$F$}

\put(203,57){\line(-4,5){20}}
\put(203,57){\line(4,3){33}}

\put(162, 48){$((F \mlc H) \mld G) \mlc F$}

\put(160, 188){sequent-calculus proof}

\put(90,0){\bf Figure 2}

\end{picture}
\end{center}

Taking  a bottom-up (conclusion-to-premises) view of proofs, the sequent-calculus proof starts with  the sequent consisting of just the target formula. On the left, at the same level, we wee the corresponding target cirquent with a single group. The number of leaves in the sequent-calculus proof tree is thus 1, and so is the number of groups at the corresponding step of the cirquent-calculus proof. This situation will persist throughout the rest of steps.  Due to the main connective $\mlc$ in $((F \mlc H) \mld G) \mlc F$, in the sequent-calculus proof tree the target sequent splits into two sequents $\langle (F\mlc H)\mld G\rangle$ and $\langle F\rangle$  when proceeding to the next (second from bottom) level of the proof. In the sequent-calculus proof, correspondingly, the single group of the raget cirquent splits into two groups, one containing $(F\mlc H)\mld G$ and the other $F$.  The next step in the sequent-calculus proof replaces $\mld$ with a comma in $(F\mlc H)\mld G$. In the cirquent-calculus proof, the single formula $(F\mlc H)\mld G$ of the left group correspondingly splits into $F\mlc H$ and $G$. Nothing remarkable has happened so far. It is when transitioning to the final (top) level of either proof that things get interesting. Due to the conjunctive formula $F\mlc H$, the sequent $\langle F\mlc H,G\rangle$ and the corresponding group of the corresponding cirquent split into two, one splinter   taking the formula $F$ with it and the other taking $H$. 
But how about $G$, which splinter can  inherit it? 
Classical logic says that both of the corresponding branches of the sequent-calculus proof tree can simultaneously take $G$, because there is no difference between multiple $G$s and a single $G$. On the other hand, the resource-conscious linear logic insists that only one of the branches can  take $G$ with it. Cirquent calculus resolves this conflict by letting $G$  be inherited by both splinter groups but explicitly showing that $G$ is shared between them, so it is still a single rather than two copies of $G$.

\section{The rules of system \textbf{CL5}}
The following theorem, proven in \cite{Cirq}, establishes the adequacy of system  \textbf{CL5}, to completing whose description the present section is devoted: 
\begin{theorem} A cirquent (of the language of \textbf{CL5}) is valid iff it has a proof in \textbf{CL5}. \end{theorem}

What follows is an informal explanations of the inference rules of \textbf{CL5}, six in total. All non-axiom rules take a single premise, and a {\bf proof} (in \textbf{CL5}) of a cirquent $C$ is undesrstood as a sequence of cirquents ending in $C$ where the first cirquent is an axiom, and every subsequent cirquent follows from its preceding one by one of the rules of inference. 

\subsection{Axioms ({\bf A})} Axioms can be thought of  as ``rules" with no premises. Every axiom of \textbf{CL5} looks like an array of groups, as shown in the figure below, where each group contains some formula and the negation of that formula. The letter ``A" next to the horizontal line is to indicate that the conclusion follows by the rule whose abbreviated name is A. Similarly for the other rules.

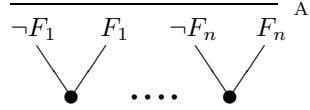
\begin{center} \begin{picture}(98,44)

\put(104,31){\scriptsize A}
\put(-3,35){\line(1,0){101}}
\put(-3,23){$\neg F_1$}
\put(31,23){$F_1$}
\put(57,23){$\neg F_n$}
\put(90,23){$F_n$}

\put(44,0){\circle*{2}}
\put(49,0){\circle*{2}}
\put(54,0){\circle*{2}}
\put(59,0){\circle*{2}}

\put(20,0){\line(-2,3){13}}
\put(20,0){\line(2,3){13}}

\put(80,0){\line(-2,3){13}}
\put(80,0){\line(2,3){13}}
\put(20,0){\circle*{5}}

\put(80,0){\circle*{5}}

\end{picture}
\end{center}

\subsection{Exchange (E)} This rule swaps arbitrary two adjacent formulas or groups, as in the two examples below. 

\begin{center} \begin{picture}(182,130)
\put(0,109){\em formula exchange}
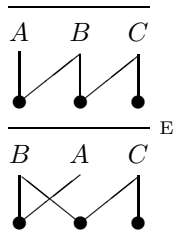
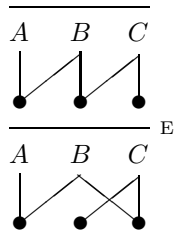
\put(10,92){\line(1,0){53}}
\put(10,79){$A$}
\put(33,79){$B$}
\put(55,79){$C$}
\put(14,56){\line(0,1){19}}
\put(14,56){\line(5,4){23}}
\put(14,56){\circle*{5}}
\put(37,56){\circle*{5}}
\put(59,56){\circle*{5}}
\put(37,56){\line(0,1){18}}
\put(37,56){\line(5,4){22}}
\put(59,56){\line(0,1){18}}

\put(10,46){\line(1,0){53}}
\put(67,44){\scriptsize E}
\put(10,33){$B$}
\put(33,33){$A$}
\put(55,33){$C$}
\put(14,10){\line(0,1){18}}
\put(14,10){\line(5,4){23}}
\put(14,10){\circle*{5}}
\put(37,10){\circle*{5}}
\put(59,10){\circle*{5}}
\put(37,10){\line(-5,4){22}}
\put(37,10){\line(5,4){22}}
\put(59,10){\line(0,1){18}}

\put(123,109){\em group exchange}
\put(130,92){\line(1,0){53}}
\put(130,79){$A$}
\put(153,79){$B$}
\put(175,79){$C$}
\put(134,56){\line(0,1){19}}
\put(134,56){\line(5,4){23}}
\put(134,56){\circle*{5}}
\put(157,56){\circle*{5}}
\put(179,56){\circle*{5}}
\put(157,56){\line(0,1){19}}
\put(157,56){\line(5,4){22}}
\put(179,56){\line(0,1){18}}

\put(130,46){\line(1,0){53}}
\put(187,44){\scriptsize E}
\put(130,33){$A$}
\put(153,33){$B$}
\put(175,33){$C$}
\put(134,10){\line(0,1){19}}
\put(134,10){\line(5,4){23}}
\put(134,10){\circle*{5}}
\put(157,10){\circle*{5}}
\put(179,10){\circle*{5}}
\put(179,10){\line(-5,4){23}}
\put(157,10){\line(5,4){22}}
\put(179,10){\line(0,1){18}}

\end{picture}
\end{center}

\subsection{Weakening (W)} Some rules are easier to describe by explaining how to get a conclusion from a premise, while some other rules, such as the present one, are more conveniently explained going from a conclusion to a premise. A premise of Weakening is obtained by deleting an arbitrary arc in the conclusion, as illustrated in the following two examples;  if,  as in the second example, a given formula was incident only with the deleted arc,  then such a formula should be deleted as well.  

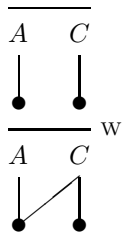
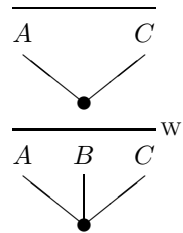
\begin{center} \begin{picture}(195,104)

\put(20,92){\line(1,0){31}}
\put(20,79){$A$}
\put(43,79){$C$}
\put(24,56){\line(0,1){18}}
\put(24,56){\circle*{5}}
\put(47,56){\circle*{5}}
\put(47,56){\line(0,1){18}}

\put(20,46){\line(1,0){31}}
\put(55,44){\scriptsize W}
\put(20,33){$A$}
\put(43,33){$C$}
\put(24,10){\line(0,1){18}}
\put(24,10){\line(5,4){23}}
\put(24,10){\circle*{5}}
\put(47,10){\circle*{5}}
\put(47,10){\line(0,1){18}}

\put(142,92){\line(1,0){54}}
\put(142,79){$A$}
\put(188,79){$C$}
\put(169,56){\circle*{5}}
\put(169,56){\line(-5,4){23}}
\put(169,56){\line(5,4){23}}

\put(142,46){\line(1,0){54}}
\put(198,44){\scriptsize W }
\put(142,33){$A$}
\put(165,33){$B$}
\put(188,33){$C$}
\put(169,10){\line(-5,4){23}}
\put(169,10){\line(5,4){23}}
\put(169,10){\line(0,1){19}}
\put(169,10){\circle*{5}}

\end{picture}
\end{center}

\subsection{Duplication (D)} This rule replaces an arbitrary group of the premise with two identical adjacent copies, as in the following example.

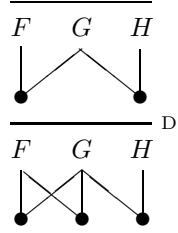
\begin{center} \begin{picture}(98,100)

\put(19,92){\line(1,0){53}}
\put(19,79){$F$}
\put(42,79){$G$}
\put(64,79){$H$}
\put(23,56){\line(0,1){19}}
\put(23,56){\line(5,4){23}}
\put(23,56){\circle*{5}}
\put(68,56){\circle*{5}}
\put(68,56){\line(0,1){18}}
\put(68,56){\line(-5,4){23}}

\put(19,46){\line(1,0){53}}
\put(76,44){\scriptsize D}
\put(19,33){$F$}
\put(42,33){$G$}
\put(64,33){$H$}
\put(23,10){\line(0,1){18}}
\put(23,10){\line(5,4){23}}
\put(23,10){\circle*{5}}
\put(46,10){\circle*{5}}
\put(69,10){\circle*{5}}
\put(46,10){\line(-5,4){22}}
\put(46,10){\line(0,1){18}}
\put(69,10){\line(0,1){18}}
\put(69,10){\line(-5,4){23}}

\end{picture}
\end{center}

\subsection{$\mld$-introduction ({\bf $\mld$})}  
To obtain a conclusion by this rule, replace any two adjacent formulas $F$ and $G$ of the premise with $ F \mld G$, and redirects it to all the arcs originally pointing to either $F$ or $G$ or both. Example:

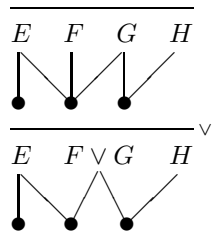
\begin{center} \begin{picture}(316,100)

\put(120,92){\line(1,0){69}}
\put(120,79){$E$}
\put(140,79){$F$}
\put(160,79){$G$}
\put(180,79){$H$}

\put(123,56){\circle*{5}}
\put(143,56){\circle*{5}}
\put(163,56){\circle*{5}}
\put(123,56){\line(0,1){19}}
\put(143,56){\line(0,1){19}}
\put(143,56){\line(-1,1){19}}
\put(143,56){\line(1,1){19}}
\put(163,56){\line(0,1){19}}
\put(163,56){\line(1,1){19}}

\put(120,46){\line(1,0){69}}
\put(191,44){\scriptsize $\mld$}
\put(120,33){$E$}
\put(140,33){$F\mld G$}
\put(180,33){$H$}
\put(123,10){\circle*{5}}
\put(143,10){\circle*{5}}
\put(164,10){\circle*{5}}
\put(123,10){\line(0,1){19}}
\put(143,10){\line(1,2){10}}
\put(143,10){\line(-1,1){19}}
\put(164,10){\line(-1,2){10}}
\put(164,10){\line(1,1){19}}

\end{picture}
\end{center}

\subsection{$\mlc$-introduction ($\mlc$)}
To obtain a premise by this rule, do the following:
 \begin{itemize}
     \item Replace a formula $F\mlc G$  of the conclusion with two adjacent formulas $ F$ and $G$.
     \item Then duplicate each group originally containing $F\mlc G$; let one copy contain $F$ and the other $G$, without otherwise changing any old containments.
 \end{itemize}
Below are two examples: 
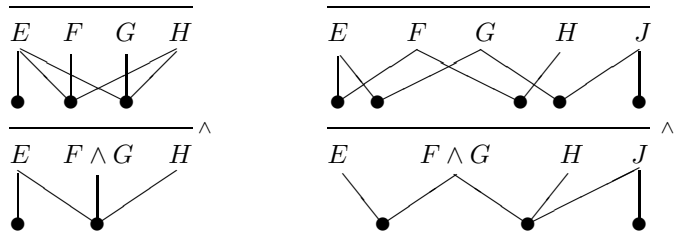
\begin{center} \begin{picture}(230,100)
\put(0,92){\line(1,0){69}}
\put(0,79){$E$}
\put(20,79){$F$}
\put(40,79){$G$}
\put(60,79){$H$}
\put(3,56){\circle*{5}}
\put(3,56){\line(0,1){19}}
\put(23,56){\circle*{5}}
\put(23,56){\line(0,1){18}}
\put(23,56){\line(-1,1){19}}
\put(23,56){\line(2,1){40}}
\put(44,56){\line(-2,1){40}}
\put(44,56){\circle*{5}}
\put(44,56){\line(0,1){18}}
\put(44,56){\line(1,1){19}}
\put(0,46){\line(1,0){69}}
\put(71,44){\scriptsize $\mlc$}
\put(0,33){$E$}
\put(20,33){$F\mlc G$}
\put(60,33){$H$}
\put(3,10){\circle*{5}}
\put(3,10){\line(0,1){19}}
\put(33,10){\circle*{5}}
\put(33,10){\line(0,1){18}}
\put(33,10){\line(-3,2){30}}
\put(33,10){\line(3,2){30}}

\put(120,92){\line(1,0){122}}
\put(120,79){$E$}
\put(150,79){$F$}
\put(176,79){$G$}
\put(206,79){$H$}
\put(236,79){$J$}
\put(139,56){\line(-4,5){14}}
\put(139,56){\line(2,1){39}}
\put(139,56){\circle*{5}}
\put(124,56){\circle*{5}}
\put(124,56){\line(0,1){17}}
\put(124,56){\line(3,2){30}}
\put(208,56){\line(-3,2){30}}
\put(193,56){\line(4,5){15}}
\put(193,56){\line(-2,1){39}}
\put(208,56){\line(3,2){30}}
\put(193,56){\circle*{5}}
\put(208,56){\circle*{5}}
\put(238,56){\circle*{5}}
\put(238,56){\line(0,1){19}}

\put(120,46){\line(1,0){122}}
\put(246,44){\scriptsize $\mlc$}

\put(120,33){$E$}
\put(155,33){$F\mlc G$}
\put(208,33){$H$}
\put(235,33){$J$}
\put(141,10){\line(-4,5){15}}
\put(141,10){\line(3,2){28}}
\put(141,10){\circle*{5}}
\put(196,10){\line(-3,2){28}}
\put(196,10){\line(4,5){15}}
\put(196,10){\line(2,1){42}}
\put(196,10){\circle*{5}}
\put(238,10){\circle*{5}}
\put(238,10){\line(0,1){20}}

\end{picture}
\end{center}
\vspace{20pt}

\section{Do not throw the baby out!}
Below is an example of a {\bf CL5}-proof. We call its target formula  
\[\bigl(( \neg P \mld \neg Q ) \mlc (\neg R \mld \neg S)\bigr) \mld \bigl((P \mld R) \mlc (Q \mld S)\bigr)\]
{\bf  Blass's principle}, as Andreas Blass  \cite{Bla92} was the first to observe its game-semantical validity. The steps marked as {\scriptsize  E...E}  and 
{\scriptsize $\mld...\mld $} are combinations of several applications of the corresponding rules (E) and $(\mld)$.

\begin{center}\begin{picture}(228,120)

\put(0,96){\line(1,0){219}}
\put(222,95){\scriptsize A}
\put(0,83){$\gneg P$}
\put(28,83){$P$}
\put(22,60){\line(-2,3){13}}
\put(22,60){\line(2,3){13}}
\put(22,60){\circle*{5}}

\put(60,83){$\gneg Q$}
\put(90,83){$Q$}
\put(82,60){\line(-2,3){13}}
\put(82,60){\line(2,3){13}}
\put(82,60){\circle*{5}}

\put(124,83){$\gneg S$}
\put(152,83){$S$}
\put(142,60){\line(-2,3){13}}
\put(142,60){\line(2,3){13}}
\put(142,60){\circle*{5}}

\put(184,83){$\gneg R$}
\put(210,83){$R$}
\put(202,60){\line(-2,3){13}}
\put(202,60){\line(2,3){13}}
\put(202,60){\circle*{5}}

\put(0,46){\line(1,0){220}}
\put(222,45){\scriptsize E...E}
\put(0,33){$\gneg P$}
\put(28,33){$\gneg Q$}
\put(22,10){\line(-2,3){13}}
\put(22,10){\line(5,1){104}}
\put(22,10){\circle*{5}}

\put(60,33){$\gneg R$}
\put(86,33){$\gneg S$}
\put(82,10){\line(-2,1){40}}
\put(82,10){\line(5,1){102}}
\put(82,10){\circle*{5}}

\put(124,33){$P$}
\put(152,33){$R$}
\put(142,10){\line(-3,1){66}}
\put(142,10){\line(2,3){13}}
\put(142,10){\circle*{5}}

\put(184,33){$Q$}
\put(210,33){$S$}
\put(202,10){\line(-5,1){106}}
\put(202,10){\line(2,3){13}}
\put(202,10){\circle*{5}}
\end{picture}
\end{center}

\begin{center}\begin{picture}(228,34)
\put(0,46){\line(1,0){220}}
\put(222,45){\scriptsize $\mld...\mld $}
\put(0,33){$\gneg P$}
\put(28,33){$\gneg Q$}
\put(19,33){$\mld$}
\put(22,10){\line(0,1){20}}
\put(22,10){\line(6,1){120}}
\put(22,10){\circle*{5}}

\put(60,33){$\gneg R$}
\put(86,33){$\gneg S$}
\put(78,33){$\mld$}
\put(82,10){\line(-3,1){59}}
\put(82,10){\circle*{5}}
\put(82,10){\line(6,1){120}}

\put(124,33){$P$}
\put(152,33){$R$}
\put(139,33){$\mld$}
\put(142,10){\line(-3,1){60}}
\put(142,10){\line(0,1){20}}
\put(142,10){\circle*{5}}

\put(184,33){$Q$}
\put(210,33){$S$}
\put(202,10){\line(-6,1){118}}
\put(202,10){\line(0,1){20}}
\put(202,10){\circle*{5}}
\put(198,33){$\mld$}
\end{picture}
\end{center}

\begin{center}\begin{picture}(228,34)
\put(0,46){\line(1,0){223}}
\put(225,45){\scriptsize $\mlc$}
\put(0,33){$\gneg P$}
\put(28,33){$\gneg Q$}
\put(19,33){$\mld$}

\put(60,33){$\gneg R$}
\put(86,33){$\gneg S$}
\put(78,33){$\mld$}
\put(82,10){\line(-3,1){59}}
\put(82,10){\circle*{5}}
\put(82,10){\line(5,1){50}}
\put(131,20){\line(4,1){41}}

\put(121,33){$(P$}
\put(151,33){$R)$}
\put(139,33){$\mld$}
\put(142,10){\line(-3,1){60}}
\put(142,10){\line(3,2){30}}
\put(142,10){\circle*{5}}

\put(169,33){$\mlc$}

\put(182,33){$(Q$}
\put(210,33){$S)$}

\put(198,33){$\mld$}
\end{picture}
\end{center}

\begin{center}\begin{picture}(228,34)
\put(-3,46){\line(1,0){226}}
\put(225,45){\scriptsize $\mlc$}
\put(-3,33){$(\gneg P$}
\put(28,33){$\gneg Q$)}
\put(19,33){$\mld$}

\put(49,33){$\mlc$}

\put(57,33){$(\gneg R$}
\put(86,33){$\gneg S)$}
\put(78,33){$\mld$}
\put(113,10){\line(-3,1){60}}
\put(113,10){\circle*{5}}
\put(113,10){\line(3,1){59}}

\put(121,33){$(P$}
\put(151,33){$R)$}
\put(139,33){$\mld$}

\put(169,33){$\mlc$}

\put(182,33){$(Q$}
\put(210,33){$S)$}

\put(198,33){$\mld$}
\end{picture}
\end{center}

\begin{center}\begin{picture}(228,11)
\put(-6,23){\line(1,0){232}}
\put(228,22){\scriptsize $\mld$}
\put(-6,10){$\bigl((\gneg P$}
\put(28,10){$\gneg Q$)}
\put(19,10){$\mld$}

\put(49,10){$\mlc$}

\put(57,10){$(\gneg R$}
\put(86,10){$\gneg S)\bigr)$}
\put(78,10){$\mld$}
\put(110,10){$\mld$}

\put(118,10){$\bigl((P$}
\put(151,10){$R)$}
\put(139,10){$\mld$}

\put(169,10){$\mlc$}

\put(182,10){$(Q$}
\put(210,10){$S)\bigr)$}

\put(198,10){$\mld$}
\put(30,-15){\textbf{Figure 3}: A proof of Blass's principle}
\end{picture}
\end{center}
\vspace{25pt}

What is interesting about Blass's principle is that, while valid in CoL and, correspondingly,  provable in {\bf CL5}, it is not provable in linear logic or even the stronger version of it known as affine logic (=linear logic with the weakening rule of sequent calculus retained). Together with obvious similarities, there are thus also discrepancies  between linear logic and CoL, even if we limit ourselves to formulas only, the common syntactic denominator of the two approaches. As both linear logic and CoL claim to be logics of resources, it is worthwhile to point out  the main cause of disagreements between them. CoL started from a mathematically strict and intuitively convincing resource semantics, and only after that tried to answer what is valid and what is not in this semantics. In CoL, we can thus meaningfully talk about  validity and prove the resource-semantical completeness of a system. As for linear logic, there is no formal resource semantics there, and the associated claims of being ``the'' logic of resources have basically remained at a level of naive examples in the ``can you get both a candy and a gum for \$1?'' style. Of course, certain formal semantics have been retroactively developed for linear logic (e.g., the phase or coherence spaces semantics), but those can hardly claim to be semantics of {\em resorces}. By and large, as a ``resource logic'',  linear logic has been conceived proof-theoretically rather than semantically, by taking classical sequent calculus and deleting rules  that seemed to be unacceptable from a naive, intuitive  resource-semantical point of view. The deleted rules were indeed wrong, such as  contraction, which is obviously inconsistent with any reasonable understanding of resources. However, where is a guarantee that, together with clearly offensive postulates, some   deeply hidden resource-semantically irreproachable derivative principles were not expelled as well? CoL believes that this is exaxtly what happened, with Blass's principle being an example of an innocent baby thrown out with the bath water.

\bibliographystyle{plain}

\begin{thebibliography}{50}

\bibitem{bauerLMCS} M. Bauer. {\em The computational complexity of propositional cirquent calculus}. {\bf Logical Methods is Computer Science 11} (1:12), 2015,  pp. 1-16.

\bibitem{Bla92} A.  Blass {\em A game semantics for linear logic}, Annals of Pure and Applied Logic, 56 (1-3) 1992, pp. 183-220.

\bibitem{anupam}  A. Das and L. Strassburger. {\em On linear rewriting systems for Boolean logic and some applications to
proof theory}. {\bf Logical Methods in Computer Science} 12 (2016), pp. 1-27.

\bibitem{Jap03} G. Japaridze. {\em Introduction to computability logic}. {\bf Annals of Pure and Applied Logic} 123 (2003), pp. 1-99.



\bibitem{Cirq} G. Japaridze. {\em Introduction to cirquent calculus and abstract resource semantics}. {\bf Journal of Logic and Computation} 16 (2006), pp. 489-532.
 

\bibitem{Japdeep} G. Japaridze. {\em Cirquent calculus deepened}. {\bf Journal of Logic and Computation}  18 (2008),  pp. 983-1028.

\bibitem{fromto} G.Japaridze. {\em From formulas to cirquents in computability logic}. 
{\bf Logical Methods is Computer Science} 7 (2011), Issue 2 , Paper 1, pp. 1-55.

\bibitem{taming1} G. Japaridze. {\em The taming of recurrences in computability logic through cirquent calculus, Part I}. {\bf Archive for Mathematical Logic} 52 (2013),  pp. 173-212.
 
\bibitem{taming2} G. Japaridze. {\em The taming of recurrences in computability logic through cirquent calculus, Part II}. {\bf Archive for Mathematical Logic} 52 (2013),  pp. 213-259.


\bibitem{cl16} G. Japaridze. {\em Elementary-base cirquent calculus I: Parallel and choice connectives}. 
{\bf Journal of Applied Logics - IfCoLoG Journal of Logics and their Applications} 5 (2018), no.1, pp. 367-388. 

\bibitem{cl17} G.Japaridze.
 {\em Elementary-base cirquent calculus II: Choice quantifiers}. 
{\bf Logic Journal of the IGPL} 2020; jzaa022, https://doi.org/10.1093/jigpal/jzaa022 .

\bibitem{fundamentals} G. Japaridze. {\em Fundamentals of computability logic 2020}. {\bf IfCoLog Jpurnal of Logics and their Applications}, 7 (2020), pp. 1177-1198.

\bibitem{XuIGPL} W. Xu and S. Liu. {\em Soundness and completeness of the cirquent calculus system CL6 for computability logic}. {\bf Logic Journal of the IGPL} 20 (2012), pp. 317-330. 
 

\bibitem{XuIf} W. Xu. {\em A propositional system induced by Japaridze's approach to IF logic}. {\bf Logic Journal of the IGPL}  22 (2014), pp. 982-991.

\bibitem{XuLast} W. Xu. {\em A cirquent calculus system with clustering and ranking}. {\bf Journal of Applied Logic}  16 (2016), pp. 37-49.



\end{thebibliography}

\end{document}